\documentclass[aps,pra,a4paper,showpacs,twocolumn]{revtex4}

\input epsf

\usepackage{amssymb}
\usepackage{amsmath}
\usepackage{amsfonts}
\usepackage{graphicx}
\usepackage{epsfig}

\def\be{\begin{equation}}
\def\ee{\end{equation}}
\def\ba{\begin{eqnarray}}
\def\ea{\end{eqnarray}}
\def\MZ{MZ}
\def\fig{Fig. }
\def\Eq{Eq. }

\begin{document}

\renewcommand{\figurename}{Figure}

\title{Phase noise measurements in long fiber interferometers for quantum repeater applications}
\author{Ji\v{r}\'{i} Min\'{a}\v{r}$^{1}$, Hugues de Riedmatten$^{1}$, Christoph Simon$^{1}$, Hugo Zbinden$^{1}$ and Nicolas Gisin$^{1}$}

\author{}
\affiliation{$^{1}$ Group of Applied Physics, University of
Geneva, CH-Geneva, Switzerland}

\begin{abstract}

Many protocols for long distance quantum communication require
interferometric phase stability over long distances of optical
fibers.  In this paper we investigate the phase noise in long
optical fibers both in laboratory environment and in installed
commercial fibers in an urban environment over short time scales
(up to hundreds of $\mu$s). We show that the phase fluctuations
during the travel time of optical pulses in long fiber loops are
small enough to obtain high visibility first order interference
fringes in a Sagnac interferometer configuration for fiber lengths
up to 75 km. We also measure phase fluctuations in a Mach-Zehnder
interferometer in installed fibers with arm length 36.5 km. We
verify that the phase noise respects gaussian distribution and
measure the mean phase change as a function of time difference.
The typical time needed for a mean phase change of 0.1 rad is of
order of 100 $\mu$s, which provides information about the time
scale available for active phase stabilization. Our results are
relevant for future implementations of quantum repeaters in
installed optical fiber networks.

\end{abstract}

\maketitle
\section{Introduction}
\label{Introduction}

Distributing quantum resources over large distances is an
important experimental challenge in quantum information science
\cite{gisin07}. One possibility to overcome losses in optical
fibers is to implement a quantum repeater architecture
\cite{Briegel98} where the total distance is split into several
segments. Entanglement between neighboring nodes in each segment
is created independently and stored in quantum memories. The
entanglement is then extended to longer distances by entanglement
swapping.

\begin{center}
\begin{figure}[h]
\epsfxsize=0.3\textwidth \epsfbox{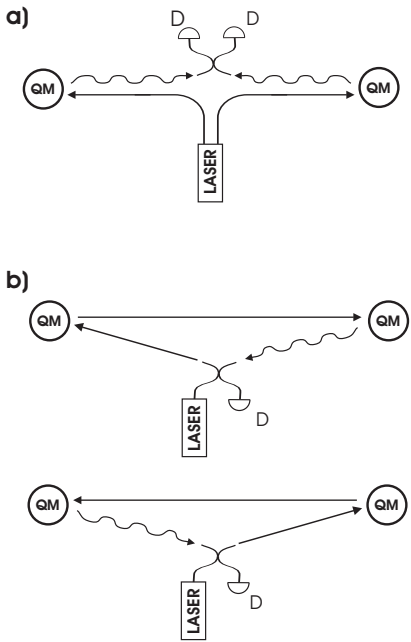} \caption{Creation of
entanglement by single photon detection. (a) The two quantum
memories (QM) are placed in the arms of a balanced Mach-Zehnder
interferometer and excited with a common laser. The emitted fields
are combined at a beam-splitter, which erases information about
the origin of the emitted photon. The detection of a single photon
with detector D after the beam-splitter projects the remote QM
into an entangled state. In order to generate entanglement,
interferometric stability must be preserved for the duration of
the experiment. (b) Sagnac configuration. The excitation pulses
for each QM are first reflected at the other QM, using e.g.
optical switches. In this way, the excitation lasers for the two
systems and the emitted photons travel the same path in a
counterpropagating way. Straight lines indicate excitation lasers
and wavy lines emitted fields. \label{QR}}
\end{figure}
\end{center}

One way to create entanglement between remote material systems is
by quantum interference in the detection of a single photon
emitted by the atomic systems
\cite{Cabrillo99,Bose99,Duan01,Childress05,Childress06,Chou05,Laurat07}
(see \fig\ref{QR}a). The remote atoms are excited in a coherent
fashion and the emitted fields are combined at a beam splitter
placed at a central location. If the two fields are
indistinguishable, the information about the origin of the photon
is erased and the detection of a single photon projects the two
atomic systems in an entangled state. This type of entanglement
creation is probabilistic, but heralded. It has been proposed both
for single systems (e.g. single atoms \cite{Cabrillo99,Bose99} or
Nitrogen-Vacancy centers in diamond
\cite{Childress05,Childress06}) and for atomic ensembles
\cite{Duan01,Chou05}. In the case of ensembles, the creation of
entanglement by single photon detection results in entangled
number states, with one joint collective atomic excitation
delocalized between the remote ensembles, as has been recently
demonstrated experimentally \cite{Chou05,Laurat07}. The advantage
of this type of entanglement is that only one photon must be
created, transmitted and detected, which enables a greater
probability per trial to obtain the desired entangled state, as
compared to the creation of entanglement by detection of two
photons \cite{Feng03,Duan03,Simon03,Chen07,Moehring07}. However,
the drawback of this method is that it requires interferometric
stability over large distances, which is generally considered to
be a challenging experimental task. Phase noise in the quantum
channels acts as a decoherence in the entanglement generation, in
the same way as atomic dephasing acts as a decoherence in the
entanglement storage process.
Several solutions have been proposed to alleviate the phase
stability problem. 1) First, the two atomic systems can be excited in
a Sagnac interferometer configuration \cite{Childress05} (see
\fig\ref{QR}b). In this way, the excitation lasers for the two memories
and the emitted photons travel the same path in a
counterpropagating fashion. Hence, as long as the phase
fluctuations are slower than the travel time, the phase difference
is automatically zero. 2) The second possibility is to stabilize
actively the phases, with respect to a reference laser. This
requires that the phase fluctuations are not too fast in order to
be able to implement a feed-back loop \cite{Chou05}. 3) Finally, a
third possible solution has been proposed in \cite{Duan01} for the
case of atomic ensembles, by implementing two chains of entangled
memories in parallel, with the same quantum channel linking them.
In that case, entanglement is generated independently in the two
chains, and stored in the memories. By selecting the case where at
least one excitation is present in each node, one can obtain an
effective two-excitations maximally entangled state. Effective
entanglement between remote quantum nodes can thus be created by
asynchronous two photon quantum interference \cite{chou07}. In
that case, the phase of the quantum channels must be constant only
during the time $\Delta t$ between the detection heralding the
entanglement in the two chains.

While there is an active theoretical activity in developing new
quantum repeaters architectures based on entanglement generation
by single photon quantum interference
\cite{Duan01,Childress05,Childress06,collins07,Simon07,Sangouard07,Jiang07},
no study has addressed so far the feasibility of the
implementation of such protocols in installed telecom networks
with respect to phase stability. In this context, we report in
this paper measurements of the phase noise in long fiber
interferometers over short timescale. Our results show that the
phase in long optical fibers (several tens of km) remains stable
at an acceptable level for times of order of 100 $\mu$s, in
realistic environments.

The measurement of phase noise in optical fibers is also relevant
for other applications. There is currently an active area of
research aiming at the transmission of frequency references over
large distances in optical fibers, in order to synchronize or
compare remote optical-frequency atomic clocks
\cite{Ma94,Coddington07,Foreman07, Foreman07b,Newbury07}. In this
case, the phase noise in the fiber link directly translates into a
spectral broadening of the frequency reference. To preserve the
precision of optical clocks, the light should be transmitted with
sub femtosecond jitter over long distances through the fiber.
Noise cancellation schemes have been developed, which work well as
long as the phase noise is negligible during the round trip time
of the fiber \cite{Foreman07b}.

In order to perform the phase noise measurements, we used two
interferometric techniques, namely Sagnac and Mach-Zehnder (\MZ)
interferometry. The Sagnac configuration allows us to study the
feasibility of solving the phase stability problem by cancelling
the phase fluctuations with the geometry of the interferometer.
Moreover it permits us to visualize the effects of phase
fluctuations directly on the visibility of first order
interference, which can then be used as a measure of the fidelity
of the quantum communication. However, with this technique we can
infer the phase noise only on a time scale shorter than the travel
time of the pulses in the interferometer. On the other hand, the
\MZ\ interferometer provides some information about the structure
of the phase fluctuations over larger time scales. This
information is important for the two other proposed techniques to
alleviate the phase noise problem, namely active stabilization and
asynchronous two-photon interference. The measurements were
performed both in spooled optical fibers the laboratory, and in
installed commercial fibers in an urban environment. The details
of the measurements are presented in sections \ref{Sagnac} and
\ref{MZ} and discussed in section \ref{Discussion}.

\section{Sagnac interferometry}
\label{Sagnac}

\begin{center}
\begin{figure}
\epsfxsize=0.48\textwidth \epsfbox{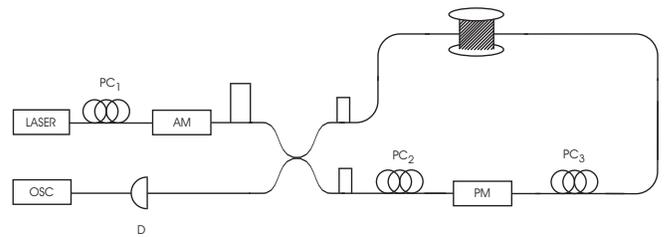}
\caption{Sagnac
interferometer: experimental setup.
A rectangular laser pulse created by amplitude modulator
AM is split at the coupler and phase is applied to one of the
pulses by the phase modulator PM. The resulting interference signal is detected by detector
D and oscilloscope OSC. PC$\mathrm{_{i}}$ denote polarization controllers.}
\label{fig Sagnac}
\end{figure}
\end{center}

Information about the phase noise of a long optical fiber can be
inferred by measuring first order interference fringes in a Sagnac
interferometer. In such an interferometer the two paths correspond
to pulses counterpropagating in the same fiber loop. For small
distances, the phase in the loop doesn't have the time to
fluctuate during the travel time of the pulses, which leads to a
zero phase difference between the two paths of the interferometer.
By changing the phase difference between the two pulses (for
instance using time resolved phase modulation) it is possible to
obtain an interference fringe. If there is no phase noise in the
interferometer during the travel time of the pulses, the
visibility will be perfect. As mentioned before, this phase
fluctuation cancellation technique can be exploited in order to
generate remote entanglement of quantum memories by single photon
interference. However, for long fibers, the phase might have time
to fluctuate during the travel time of the pulses, because the
pulses travel through a given segment of the fiber at different
times. In that case, the visibility of the interference fringe
will be reduced. Hence, one can use the visibility as a measure of
the phase noise over a  time scale shorter than the time of
propagation of the light in the interferometer. Let us now
investigate in more detail the relation between visibility and
phase noise.

Consider that the intensity $I$ at the output of the
interferometer is of the form \be
I(\delta\varphi)=\frac{I_0}{2}\left[1+\cos(\varphi+\delta\varphi)\right]
\label{eq Intensity} \ee where $I_0$ is the average intensity and
$\varphi$ is a constant phase given by the difference between the
two arms of the interferometer. In this model the instantaneous
value of intensity I is given by a particular phase fluctuation
$\delta\varphi$. Next we assume that the distribution of phase
fluctuations reads \be
p\,(\delta\varphi)=\frac{1}{\sigma\sqrt{2\pi}}\,e^{-\frac{\delta\varphi^2}{2\sigma^2}}
\label{eq Distribution} \ee where $\sigma$ is the width of the
gaussian distribution. The assumption of a gaussian distribution
of the phase noise will be justified by our results presented in
the next section. In order to obtain some characteristic phase
noise independent value $\left\langle I \right\rangle$ of
intensity , one has to average the intensity described by
\Eq(\ref{eq Intensity}), such that \be \left\langle I
\right\rangle =
\int^{\infty}_{-\infty}\,d\delta\varphi\,p(\delta\varphi)I(\delta\varphi)
=
\frac{I_0}{2}\left(1+\cos{\varphi}\,e^{-\frac{\sigma^2}{2}}\right)
\ee We can thus find a direct relation between the visibility $V$
and the distribution of the phase noise described by  $\sigma$ \be
V=e^{-\frac{\sigma^2}{2}} \label{eq Visibility} \ee

\begin{center}
\begin{figure}
\epsfxsize=0.48\textwidth \epsfbox{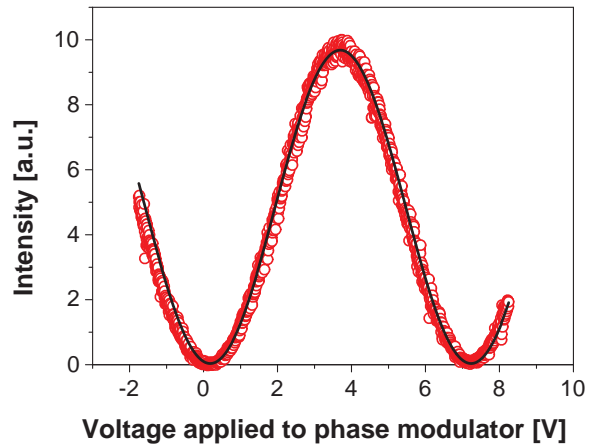}
\caption{(Color online) Example of interference fringe for a Sagnac
interferometer of 36.5 km in the telecom network. The solid curve
is a sinusoidal fit which gives the visibility 99.2\%. The time
 to scan a fringe is about 10 minutes, which shows that fast
phase fluctuations do not degrade the visibility significantly in
this case.} \label{fringe}
\end{figure}
\end{center}
Let us now describe our experimental setup, shown in \fig\ref{fig
Sagnac}. The light source is a single mode Distributed Feedback
laser diode at 1556 nm (linewidth  1 MHz). The light emitted by
the laser is modulated by an integrated LiNbO$_3$ optical
amplitude modulator (AM) which creates 1-10 $\mu$s rectangular
light pulses with repetition rate 1 kHz. These pulses are split at
the fiber beam splitter and sent to the Sagnac interferometer. We
apply the phase to only one of the pulses by gating a phase
modulator (PM) with a pulsed variable voltage. Finally, the
interfering light pulses are combined at the fiber beam splitter
and the resulting interference signal is detected using a detector
and oscilloscope. The area of the pulse is then plotted as a
function of the phase shift. The phase is scanned linearly, which
results in a sinusoidal interference fringe. The visibility of the
fringe is determined by a sinusoidal fit. All the visibilities
were obtained after the subtraction of the detector noise from the
signal. The phase in the modulator is varied slowly in order to
sample the entire gaussian distribution defined in Eq. (\ref{eq Distribution}). The typical time for scanning
one fringe is chosen to be about 10 min.

We used several polarization controllers (PC) in order to ensure
the optimal polarization alignment. At the same time we made sure
that this alignment was quite stable over tens of minutes so that
the quick intensity changes are not due to polarization
fluctuations. For the experiment in the laboratory conditions, we
used spools with different lengths. The spools were inserted in a
polystyren box in order to prevent fast thermal fluctuations. On
the other hand, for the measurement in installed fibers, our
building is directly connected to the commercial Swisscom network.
We had at our disposal two links connecting our building to two
different telecom stations in the Geneva area, about 17.5 km
and 18.25 km away from our building, respectively. Each link
was composed of two fibers, connected together at the telecom
stations. In this way, we could use two fibers loops of 35 km
and 36.5 km, respectively, going to different locations. By
connecting the two loops together at our building, one can form a 71.5~km long Sagnac interferometer.
To connect our lab to the telecom network, we used about 150 m of optical fibers.
\begin{center}
\begin{figure}
\epsfxsize=0.42\textwidth \epsfbox{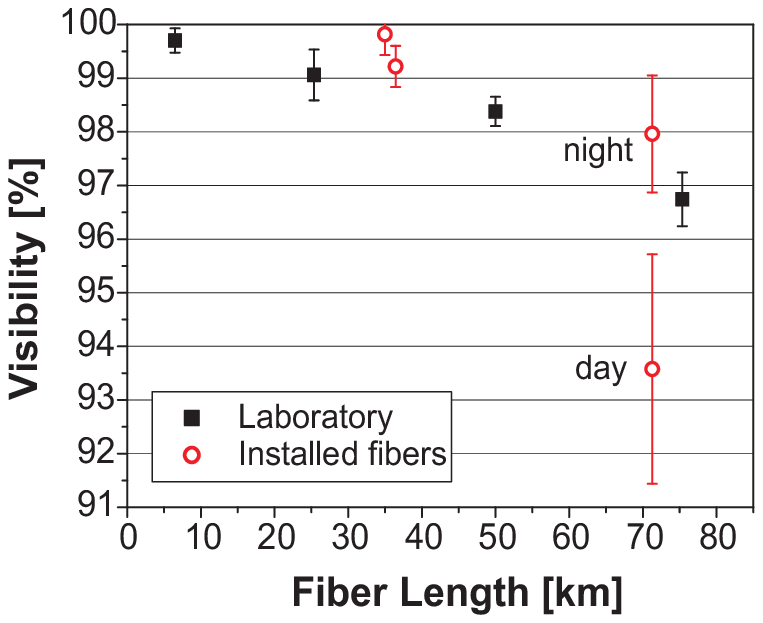}
\caption{(Color online) Visibility as a function of fiber length for the Sagnac
setup. The decrease of visibility with increasing fiber length is
obvious as well as the qualitative agreement between laboratory
and telecom measurements. High visibilities confirm the robustness
of the Sagnac interferometer setup. For the length 71.5 km,
the difference between the day (more phase noise) and night (less
phase noise) values of visibilities in the telecom network is
worth noting (the night measurements have been performed between
11 P.M. and 1 A.M.). Each point in the graph is the average of a set of
interference fringes and the errors are the standard deviations.}
\label{fig Visibility}
\end{figure}
\end{center}

A typical example of interference fringe is shown in
\fig\ref{fringe}, for the 36.5 km long Sagnac interferometer in
installed fibers.  One can see that the signal is very stable,
even for duration of order of minutes. This is important
especially in the case when one needs to accumulate sufficiently
large photon statistics. We investigated the dependence of
visibility on the fiber length both in laboratory and real world
conditions. The results are shown in \fig\ref{fig Visibility}.
For short distances, the visibility is almost perfect. As
expected, a slight decrease of the visibility can be observed when
increasing the fiber length, due to phase fluctuations during the
travel time of optical pulses. However, the visibilities remain
high even for fiber lengths  $\sim$ 75 km.  Note that in the 71.5
km interferometer in installed fibers, there is a significant
change in visibility for the measurements performed during the day
or the night. From the measured fringe visibilities, and
attributing all the loss of visibility to phase noise, we are able
to express an upper bound for the mean phase change for a given
distance using \Eq(\ref{eq Visibility}) . For instance, for the
maximal distance in installed fibers $L=71.5$ km, the measured
visibility during the day $V=93.6\%$ corresponds to $\sigma=0.36$
rad and the measured visibility during the night $V=98\%$
corresponds to $\sigma=0.2$ rad for a travel time 360 $\mu$s. One
can see that the measurements performed in installed fibers in the
urban environment give similar visibilities than those in the
laboratory (within the error). The cases where the visibilities in
the installed fibers are better than visibilities expected in the
laboratory (25 km interferometer in the laboratory and the 35
and 36.5 km interferometer in the telecom network) might be
explained by global versus local disturbance: the long fibers used
in laboratory were on spools and when some disturbance like a
mechanical vibration appeared, it was applied simultaneously to
the whole length of the fiber whereas in the telecom network such
disturbance applied only to the local part of fiber.

\section{Mach-Zehnder interferometry}
\label{MZ}

\begin{center}
\begin{figure}
\epsfxsize=0.48\textwidth \epsfbox{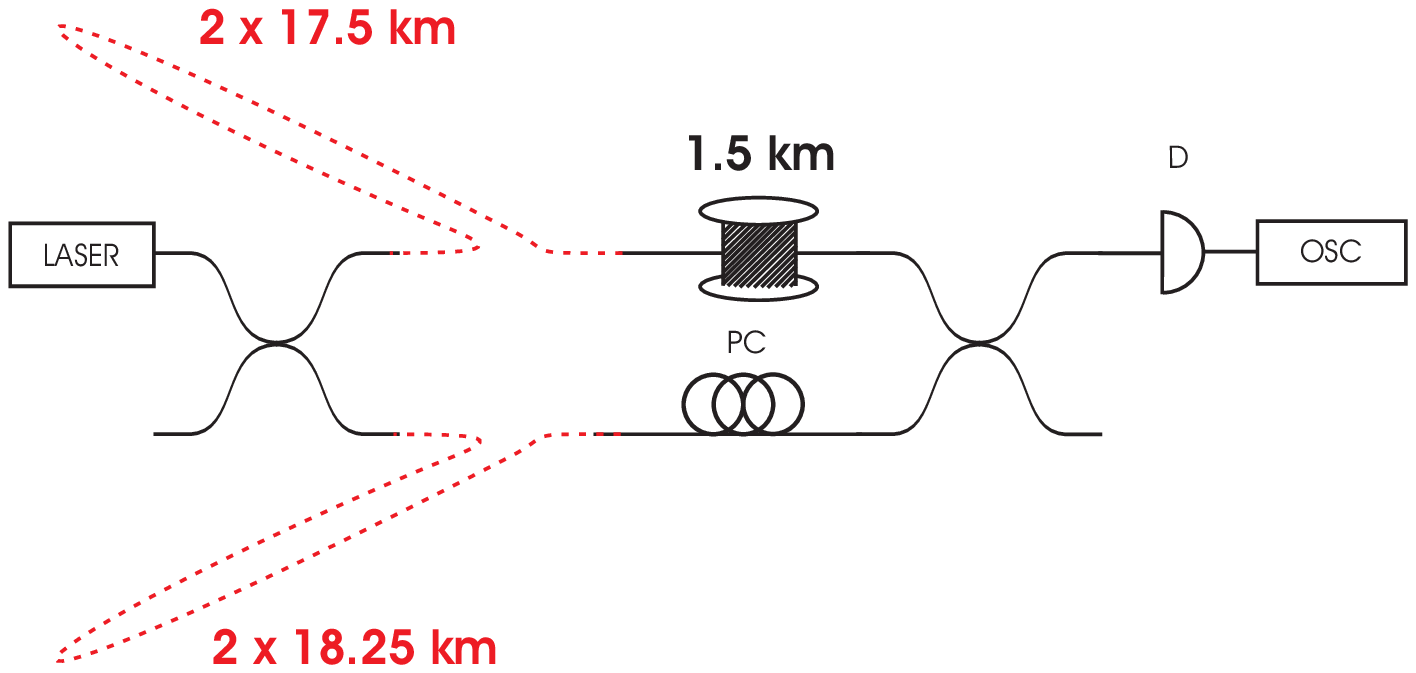} \caption{(Color online) Mach-Zehnder
interferometer: experimental setup. Dashed lines denote the
telecom fibers. The length difference between the two branches of
interferometer was few cm, i.e. much less than the coherence
length of the laser. We used the polarization controller PC$_1$ in
order to optimize the signal at the output. The resulting
interference signal is detected by detector D and oscilloscope
OSC.} \label{fig MZ}
\end{figure}
\end{center}

\begin{center}
\begin{figure}
\epsfxsize=0.48\textwidth \epsfbox{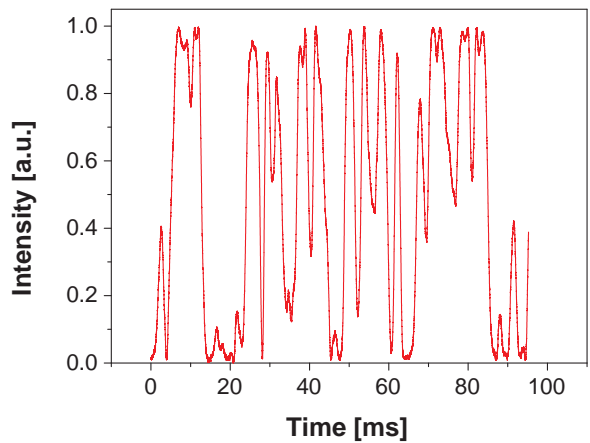} \caption{(Color online) An example
of raw measurement of intensity variation as a function of time
for the 36.5 km Mach-Zehnder interferometer in the telecom
network.} \label{fig MZ_raw}
\end{figure}
\end{center}

As mentioned above, Sagnac interferometry allows us only to obtain
information about the integrated phase noise over a time-scale
shorter than the travel time of a light pulse in the
interferometer. In order to obtain a more quantitative analysis
over longer time scales, we used Mach-Zehnder interferometry. The
experimental setup is shown in \fig\ref{fig MZ}. The two arms of
the interferometer are composed by the two fiber loops connecting
our lab to the two telecom stations. In order to balance the
lengths of the two arms, we had to add a spool of 1.5 km of fiber
in our lab. We introduced a polarization controller into one arm
of the interferometer in order to align the polarization at the
output coupler. Similarly to the experiment with the Sagnac
interferometer, we verified that there are no quick intensity
changes due to polarization fluctuations. The resulting
interference signal is detected by a detector and an oscilloscope
with  1~GHz  and 6~GHz bandwidth respectively. We verified that
there are no significant changes of light intensity at the
detector in the time scale of few $\mu$s. It is thus sufficient to
use a few datapoints to describe the intensity changes for this
time difference without loss of information. In our case the
typical distance between two adjacent datapoints was 2-4 $\mu$s,
which is then the final temporal resolution of the measurement. We
also assured that during measurement in this setup the lengths of
both arms of the interferometer were equilibrated in such a way
that their difference was of order of centimeters. This is much
less than the coherence length of the laser (66 m in the optical
fiber) corresponding to the measured bandwidth 1 MHz, and leads to
negligible phase noise induced by the frequency drift of the
laser. In this section we will focus only on measurements
performed in the installed fibers.
If we consider two interfering light fields, we can find the
dependence of measured intensity $I$ on time $t$ in the form \be
\label{I}
I(t)=\frac{I_{max}-I_{min}}{2}\left[1+\cos\varphi(t)\right]+I_{min} \ee
where $I_{max}$ and $I_{min}$ are maximal and minimal measured
intensity and $\varphi$ is the actual phase difference between the
two fields. For our purposes we call from now on this phase
difference simply the phase. An example of such a measurement is
shown in \fig\ref{fig MZ_raw}. From the measurement of $I(t)$ and
using \Eq(\ref{I}) we can directly find the time dependence of
the phase. However, around maxima and minima of the signal, a
relatively large change in the phase causes only a small change in
the intensity and thus introduces a bigger error to the data
analysis. On the other hand, on the slope of a measured signal, a
relatively small change of phase causes significant change in the
intensity. For this reason we restricted our analysis only to this
region, omitting maxima and minima; this provides us with less
data but shouldn't qualitatively change obtained statistics. We
also verified that the interference isn't caused by
retroreflections in the interferometer.

\begin{center}
\begin{figure}
\epsfxsize=0.44\textwidth \epsfbox{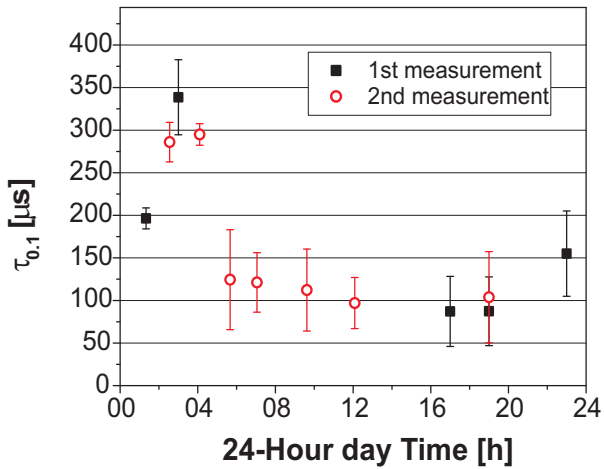} \caption{(Color online) Time
intervals needed for phase change 0.1 rad in the 36.5 km
Mach-Zehnder interferometer in the telecom network. The first set
of measurements consists of points which were measured in the span
of 12 days. The second set was obtained during a single day. All
measurements are compatible and show the signal with less phase noise during the
night (around 3 A.M.) than during the day.} \label{fig Day time}
\end{figure}
\end{center}

Once we calculated the temporal evolution of the phase, we can
investigate the phase changes as a function of time. For a time
difference $\tau$ we can find a set of corresponding phase
differences $\{\delta\varphi_\tau^j\}$, where
$\delta\varphi_\tau^j = \left| \varphi(t_{j}+\tau)-\varphi(t_{j})
\right|$, and the final phase difference as the average over $j$,
schematically: \be \tau :
\{\delta\varphi_\tau^1,\delta\varphi_\tau^2,\dots\} \rightarrow
\Delta\varphi_{\tau} \label{eq d_phis} \ee
An example of dependence $\Delta\varphi$ vs. $\tau$ is shown in \fig\ref{fig
MZ_d_phi}. Actually every measurement will give a slightly different (but still monotonously rising)curve, i.e. the times needed for, e.g., a phase change of 0.1 rad will be different. The different measurements are then averaged and statistical errors can be calculated as shown later in \fig\ref{fig Day time}. Another interesting information one can obtain from such a set of phase differences is their distribution. The
distribution for a time difference corresponding to the travel
time in the interferometer ($\tau = 182\ \mu$s) is shown in \fig\ref{fig Dist}. In the data processing to obtain this curve, we
assume that long terms phase drifts are negligible on this time
scale and that the phase fluctuations are random. Thus a positive
(negative) intensity change is attributed to a positive (negative)
phase change. Hence, the $\delta \varphi$ are considered here
without absolute value. The gaussian fit in \fig\ref{fig Dist}
shows that the phase noise distribution corresponds well to a
gaussian distribution. This can be explained if we describe the
phase fluctuations in terms of random walk theory \cite{Feller}.
\begin{center}
\begin{figure}
\epsfxsize=0.4\textwidth \epsfbox{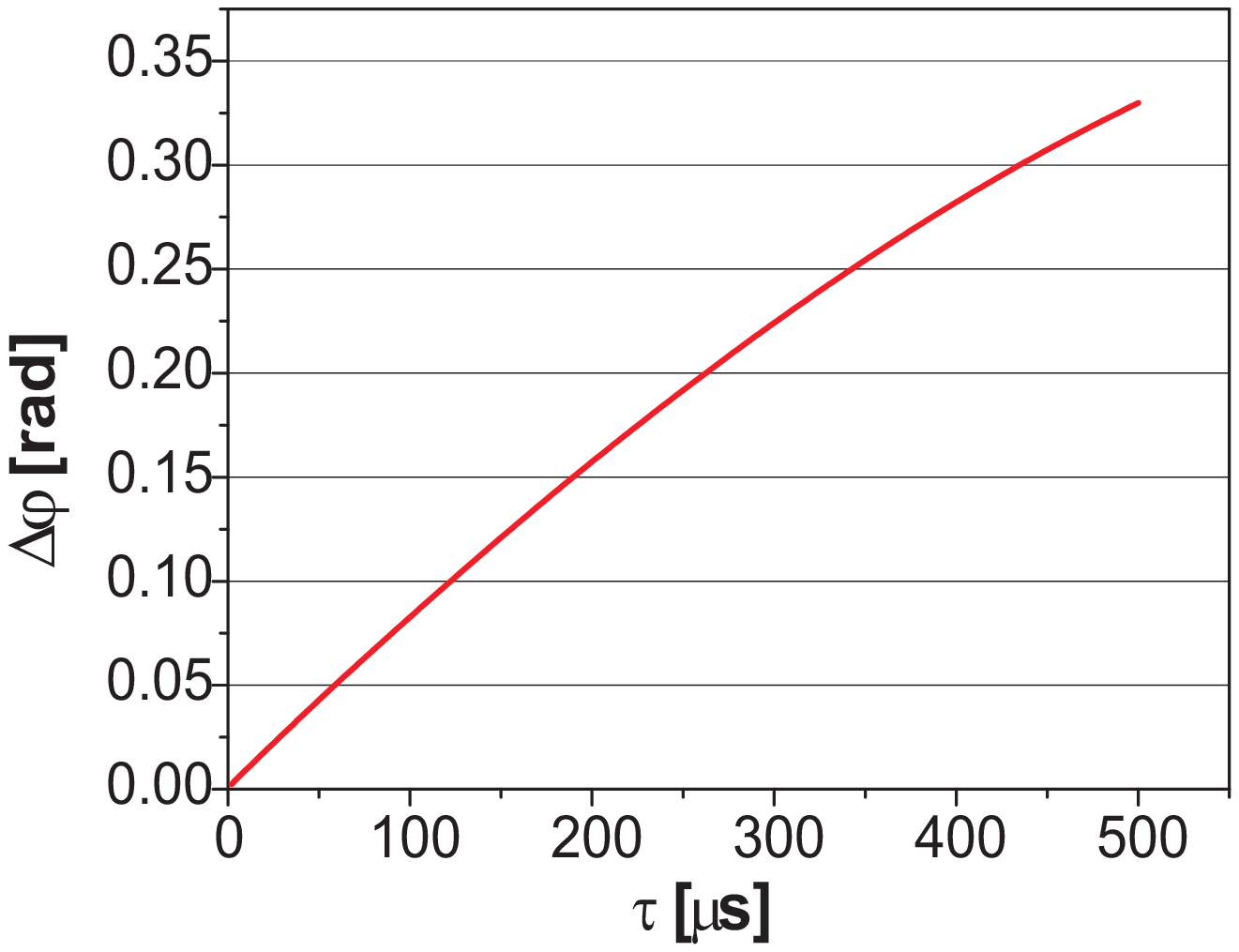}
\caption{(Color online) Dependence of the mean phase change $\Delta\varphi$,
resulting from the gaussian distribution (see \fig\ref{fig Dist}),
on time difference $\tau$ as described in \Eq(\ref{eq d_phis}). An
example for 36.5 km Mach-Zehnder interferometer in the
telecom network is shown. Here the time corresponding to the phase
change $\Delta \varphi = 0.1$ rad is $\tau_{0.1} = 122\ \mu$s.} \label{fig MZ_d_phi}
\end{figure}
\end{center}

Note that the gaussian distribution (\Eq(\ref{eq Distribution}))
describes fully the phase noise at a given time $\tau$. It is
possible to show that there is a direct relationship between this
distribution and $\Delta \varphi$ calculated in \Eq(\ref{eq
d_phis}), such that \be \Delta\varphi = \left\langle \left|
\delta\varphi \right|\right\rangle = \sqrt{\frac{2}{\pi}}\,\sigma
\label{eq d_p} \ee

\begin{center}
\begin{figure}
\epsfxsize=0.4\textwidth \epsfbox{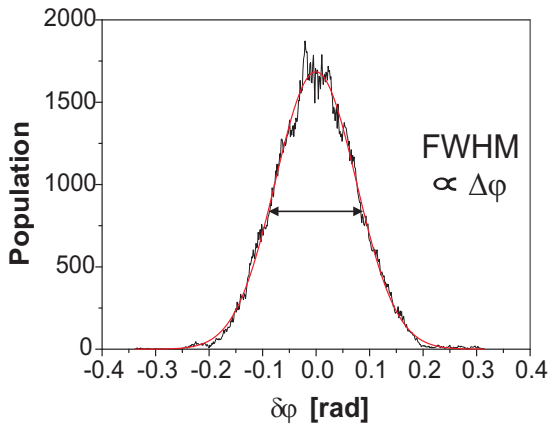} \caption{(Color online) The
distribution of the phase noise in the 36.5 km Mach-Zehnder
interferometer in the telecom network. The distribution is shown
for the time of propagation, i.e. $\tau = 182\ \mu$s. One can see that
the fit (gaussian distribution \Eq(\ref{eq Distribution}))
corresponds well to the observed results. The average phase change $\Delta\varphi$, plotted in \fig\ref{fig MZ_d_phi},
is proportional to FWHM of the phase noise distribution.}
\label{fig Dist}
\end{figure}
\end{center}
One of the possibilities to quantify the quick phase changes is to
fix some value of the phase difference, which can still be
tolerated in the mentioned quantum communication protocols and
look what is the corresponding time needed for such phase change.
We took for example the value of $\Delta\varphi=0.1$ rad, which
according to \Eq(\ref{eq Visibility}) corresponds to a visibility
of 99.5$\%$ and we found the corresponding time intervals
$\tau_{0.1}$. This value will be further justified in the
following section. We investigated how these time intervals
changed in installed fibers as a function of the time of day. The
results are plotted in \fig\ref{fig Day time}. It is obvious that
the phase is more disturbed during the day, than during the night,
with $\tau_{0.1}$ moving from $\sim$100 $\mu$s during the day to
$\sim$ 350 $\mu$s during the night. This suggests that a big part
of the induced phase noise is due to vibration caused by external
disturbances, e.g. traffic.

\section{Discussion}
\label{Discussion}

In this section, we discuss in more detail the results obtained
with the two interferometric methods. A first interesting point to
consider is the dependance of the measured phase noise on fiber
length. A common assumption is that the variance of the phase
noise $\textup{var}(\delta\varphi)$ is proportional to the length
of fiber $L$, where the constant of proportionality is the
diffusion coefficient \cite{Jiang07}. \be \textup{var}(\delta\varphi) = D\,L \ee
For the measurements in the Sagnac configuration, we can express
the diffusion coefficient as a function of the width $\sigma$ and
using \Eq(\ref{eq Visibility}) as a function of the visibility
$V$. \be D = \frac{\textup{var}(\delta\varphi)}{L} =
\sigma^2\,\frac{1}{L} =
-2\,ln\,(V)\,\frac{1}{L} \label{eq D} \ee

\begin{center}
\begin{figure}
\epsfxsize=0.42\textwidth \epsfbox{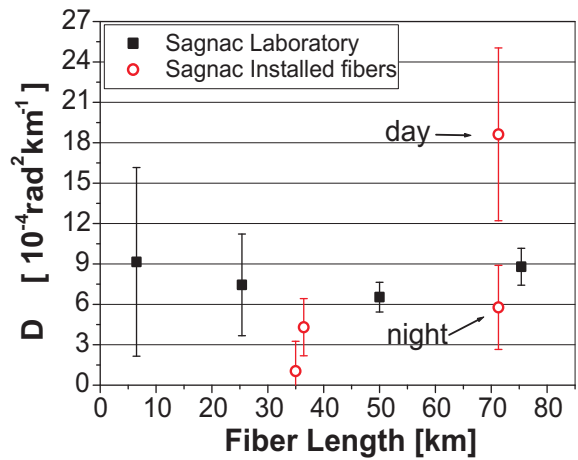}
\caption{(Color online) Diffusion coefficients in the laboratory and telecom
network. The diffusion coefficients measured in the laboratory
yield all the same value (within the error) as expected from
theory. For the 71.5 km interferometer in the telecom network, there is however a significant difference between the
measurements during the day (more phase noise, i.e. larger
coefficients) and night.} \label{fig D}
\end{figure}
\end{center}

\fig\ref{fig D} shows the calculated diffusion coefficient as a
function of fiber length both for spooled fibers in laboratory and
for installed fibers.  We observe that different fiber lengths
yield the similar diffusion coefficient (of order of $D=8
\cdot10^{-4}\ \mathrm{rad^2\,km^{-1}}$). This confirms the assumption that
the variance of phase noise is proportional to the fiber length.
As expected from \fig\ref{fig Visibility}, there is however a
difference for the value of $D$ obtained for the 71.5 km Sagnac
interferometer in installed fibers during the day and during the
night. This differences might be due to two factors. (1) The phase
noise is larger during the day than during the night, as shown in
\fig\ref{fig Day time}. (2) In this analysis, all the loss of
visibility is attributed to the phase noise, such that the
coefficient D is an upper bound. For the 71.5 km interferometer
however, the alignment of polarization was difficult due to a
noisy signal. In this case, it is thus likely that a part of the
visibility loss is also caused by a polarization mismatch. Note
that we do not observe a significant difference between day and
night for the Sagnacs interferometers of lengths 35 km and 36.5
km. This might be due to the fact that in this case the whole
fiber is in the same cable.

Let us now discuss the results obtained with the Mach-Zehnder
interferometer. If we consider the theory of random walk to
explain the structure of phase noise, we should obtain a
histogram, i.e. a distribution of phase noise in the form of a
gaussian, see \fig\ref{fig Dist}. At this point the theory
corresponds well to the observed results. The other conclusion
arising from random walk theory is that the function plotted in
\fig\ref{fig MZ_d_phi} should depend on $\tau$ as $\tau^{x}$
where $x=\frac{1}{2}$. In our case, we find value of $x$ ranging
from 0.7 to 0.9. At this point, we do not have a model for the
phase fluctuations that explains the observed dependance.

We also considered the effect of a potential slow thermal drift on
the observed phase noise. A phase shift of 0.1 rad corresponds to
an optical path length difference between the two arms of the
interferometer of about 25 nm for a wavelength of 1550 nm.
Assuming a drift constant in time (for instance due to a slow
temperature variation), an optical path difference of 25 nm in 100
$\mu$s would lead to an optical path difference of 90 cm per
hour. We measured the optical lengths of the two fibers with a
high resolution photon counting optical time domain reflectometer \cite{Wegmuller04},
and found that the difference remained constant within a few cm
over the course of 12 hours (from 1 P.M. to 1 A.M.).  This shows that
the observed phase noise is most likely due to other causes than a
slow thermal drift.

Finally, let us compare the results obtained with the two methods.
From the phase noise $\Delta \varphi (\tau=L_{MZ}/c)$ obtained
with a Mach-Zehnder interferometer with arm length $L_{MZ}$, one
can predict the visibility one should obtain with a Sagnac
interferometer of length $2L_{MZ}$, with the help of \Eq(\ref{eq
Visibility}) and \Eq(\ref{eq d_p}).  In our case the visibility
calculated from the Mach-Zehnder interferometer of $L_{MZ} = 36.5$ km is $V = (99.0 \pm 0.7)\,\%$ during the night (between 11 P.M. and 1 A.M.)  and $V = (97.1 \pm 2.4)\,\%$ during the day. This is
compatible (within the error) with the results for the Sagnac
interferometer of 71.5 km shown in \fig\ref{fig Visibility}.

\section{Applications to quantum repeaters}
Let us now analyze our results with respect to applications in
quantum repeaters. As mentioned in the introduction, there are
several ways to alleviate the phase stability problems. The Sagnac
configuration presented in section \ref{Sagnac} is very similar to the
self-aligned configuration proposed in \cite{Childress05}. In
this setup, the two ensembles are excited by a laser pulse
travelling in opposite direction in the same fiber loop. In this
way, the optical path lengths for the two excitation/emission paths
are identical, as long as the phase of the loop is stable for a
duration equal to the travel time. Our measurement in urban
environment shows that first order interferences with high
visibilities can be obtained in such a configuration, for fiber
lengths up to ~71.5 km (corresponding to a maximal physical distance
between the two nodes of ~35 km and 36.5 km). This shows that from the point
of view of phase stability, it is feasible to generate
entanglement by single photon quantum interference between two
quantum memories separated by tens of km, without active
stabilization. Moreover, by using the measurements of the
diffusion coefficients (during the night) presented in \fig\ref{fig D} and
\Eq(\ref{eq D}), visibilities higher than 90$\%$ can be inferred
for fiber lengths of order 250 km.

Alternatively, it is also possible to generate entanglement by
single photon detection with a Mach-Zehnder configuration. In that
case however the phase must be stable for the duration of the
experiment, i.e. for a duration several orders of magnitude longer
than the typical time of phase fluctuations. Hence, in this
configuration it is necessary to stabilize the phase actively. Our
measurement shows that the phase remains constant at an acceptable
level for quantum communication purposes (see below) for duration
of order of 100 $\mu$s so that the phase noise at frequencies
higher than a few tens of kHz can be neglected. Active
stabilization at this frequency range seems within reach of
current technology (although it remains experimentally
challenging). There is currently an active experimental effort to
transmit phase references over long distances in optical fibers.
In this context, the distribution of a phase reference with sub
femtosecond jitter over an actively stabilized fiber link of
 32 km length has been recently demonstrated \cite{Foreman07}.

As mentioned in the introduction, in order to perform quantum
protocols with entangled number states involving one delocalized
excitation, one possibility is to implement two chains of
entangled memories in parallel and to obtain an effectively
entangled state (with two excitations) by post-selection in the
final stage \cite{chou07}. In that case the relevant time scale
for phase stability is the time between successful entanglement
generations in the two chains. For standard quantum repeater
protocols based on asynchronous two photon interference, $\Delta
t$ can be quite long (orders of seconds), mainly because of the
small probability of success per trial and of the limited
repetition rate due to the communication time \cite{Simon07}. This
means that the phase of the quantum channels should remain stable
during this time which is four orders of magnitude longer than the
typical phase fluctuation time measured in our experiment (100
$\mu$s). Hence in that case, either a self-aligned configuration
or an active stabilization are required. However, it is in
principle possible to increase by several orders of magnitude the
entanglement generation rate with spatial, spectral or temporal
multiplexing. For example, in \cite{Simon07}, it was suggested
that the use of quantum memories allowing the storage of multiple
temporal modes (multimode memories) could decrease the required
time $\Delta t$ in order to generate successful entanglement
between remote quantum nodes from seconds to tens of microseconds.
Our measurement shows that the phase fluctuations on this time
scale are acceptable.

Let us finally evaluate how the phase noise propagates in a
quantum repeater architecture and estimate the fidelity that might
be obtained with the measured phase noise. In the following
analysis for the sake of simplicity we consider only the errors
caused by the phase noise. For one elementary link of quantum
repeater the ideal state of two entangled memories can be written
in the form \cite{Simon07} \be \left| \psi_{id} \right\rangle =
\frac{1}{\sqrt{2}}\,\left(\left|0 1\right\rangle +
e^{i\Phi}\left|1 0\right\rangle \right) \label{eq psi id} \ee
Where $\Phi$ is the phase difference in between the two arms of
the interferometer. If we consider the phase noise in optical
fibers, the state for a particular phase shift $\delta \varphi$
is: \be \left| \psi \right\rangle =
\frac{1}{\sqrt{2}}\,\left(\left|0 1\right\rangle +
e^{i(\Phi+\delta\varphi)}\,\left|1 0\right\rangle \right)
\label{eq psi real} \ee For phase shifts $\delta \varphi$
distributed with Gaussian distribution given by \Eq(\ref{eq
Distribution}), the state of the entangled memories is now given
by the density matrix $\hat{\rho}_{real}$ \be \hat{\rho}_{real} =
\int d\delta\varphi\, p\,(\delta\varphi) \left| \psi \right\rangle
\left\langle \psi \right| \ee Fidelity can then be calculated from
its definition \be F = \left\langle \psi_{id} \right|
\hat{\rho}_{real} \left| \psi_{id} \right\rangle =
\frac{1}{2}\,\left(1+e^{-\frac{\sigma^2}{2}}\right) \label{eq
Fidelity} \ee The important thing to point out is that if we
consider not one but $N$ elementary links and the corresponding
entanglement connections \cite{Duan01}, the phase shift in \Eq(\ref{eq psi real}) becomes $\delta\varphi=\delta\varphi_1 +
\delta\varphi_2 + \dots + \delta\varphi_N$ and factor $\sigma$ in
\Eq(\ref{eq Fidelity}) for fidelity becomes $\sigma^2 =
\sigma_1^2 + \sigma_2^2 + \dots + \sigma_N^2$. This is the way the
errors add in the considered protocol.

As an example we consider two distant locations separated by 1000
km, connected with 8 elementary links \cite{Duan01,Simon07}. We
fixed a desired fidelity $F=0.9$ for the entangled state between
the two distant places. The fidelity can be also expressed in
terms of observed visibility as $F=(1+V)/2$. Using \Eq(\ref{eq
Visibility}) and the fact that $\sigma^2\sim L$, we can calculate
the allowed mean phase change $\Delta \varphi_{lim}$ for the
36.5 km segment of fiber. We find $\Delta \varphi_{lim} = 0.1$ rad, which
motivated our choice of this value in section \ref{MZ}. Note that
for experimental realizations, other factors will also contribute
to the decrease of fidelity, such as dark counts, probability of
generating multiple photons, distinguishability between photons
and memory errors.

\section{Conclusion}
\label{Conclusion} The creation of entanglement between remote
quantum memories by single photon interference is attractive for
quantum repeaters applications because it is much less sensitive
to loss than two-photon schemes, and thus it increases the
probability per trial to generate the desired entangled state.
This in turns results in shorter times to distribute entanglement
over long distances. However, the need for interferometric phase
stability over long distances is generally considered problematic
from an experimental point of view. In this context, we presented
an experimental investigation of phase fluctuations over short
time scales in long fiber interferometers. Our results show that
the phase remains stable at an acceptable level for quantum
communication protocols for durations of order 100 $\mu$s for a
36.5 km long Mach-Zehnder interferometer in installed telecom
fibers. This typical time for phase fluctuation should allow an
active stabilization of the phase for long interferometers. We
also showed experimentally that the phase fluctuations in
installed fibers are slow enough to guarantee high visibility
first order interference for Sagnac interferometers over length
of several tens of km, without any active stabilization. This
demonstrates the feasibility of observing single photon
interference over these lengths, and of its use to generate
entanglement between remote quantum memories without active phase
stabilization. Our measurements have been performed in installed
fibers in a noisy urban environment. An increase in stability is
thus expected when using underwater fibers.

We would like to thank M. Legr\'{e}, D. Stucki, M.Afzelius and
C.W. Chou for useful discussions. We thank Swisscom for giving us
access to their optical fiber network. Financial support by the
Swiss NCCR Quantum Photonics and by the EU Integrated Project
Qubit Applications (QAP) is acknowledged.

\end{document}